\documentclass{article}
\usepackage{spconf,amsmath,graphicx,hyperref}
\usepackage{amssymb}
\usepackage{amsmath}
\usepackage{placeins}
\usepackage[safe]{tipa}

\title{Anatomy-Aware Cross-Speaker Adaptation of Complete Vocal-Tract Acoustic-to-Articulatory Inversion}
\name{Nhat-Nam Nguyen\textsuperscript{1} -
Pierre-André Vuissoz\textsuperscript{2} -
Yves Laprie\textsuperscript{1}
}
\address{$^{1}$Université de Lorraine, CNRS, Inria, F-54000 Nancy, France\\
$^{2}$Université de Lorraine, Inserm, IADI U1254, F-54000 Nancy, France\\
\texttt{nhat-nam.nguyen@loria.fr, pa.vuissoz@chru-nancy.fr, yves.laprie@loria.fr}}
\begin{document}
%
\maketitle
\begin{abstract}
Cross-speaker acoustic-to-articulatory inversion requires accounting for anatomical differences between speakers. We propose a geometric adaptation framework that uses anatomical landmarks, primarily on vertebrae and dental structures, to transfer predictions from a fixed inversion model to unseen speakers. An affine transformation followed by thin-plate spline (TPS) deformation maps the predicted contours of 10 vocal-tract structures into each target speaker’s geometry without retraining. Landmarks are identified in one selected /u/ frame per speaker as a common phonetic reference without assuming identical articulatory configurations across speakers, and the resulting mapping is reused across recordings. We train the model on a single-speaker rt-MRI database and evaluate adaptation on eight speakers from a separate multi-speaker rt-MRI database. We compare affine and TPS configurations using 12 or 14 landmarks. Affine12+TPS14 achieves the lowest mean point-to-closest-point
error of 3.19 mm. These results support the combined value of anatomical landmark information and nonrigid alignment.
\end{abstract}
\begin{keywords}
Acoustic-to-articulatory inversion, speaker adaptation, anatomical landmarks, thin-plate splines
\end{keywords}
\section{Introduction}
\label{sec:intro}

Acoustic-to-articulatory inversion (AAI) estimates speech
articulation from acoustics. Generalization to unseen speakers remains challenging because vocal-tract anatomy and articulatory strategies vary across individuals \cite{serrurier2019characterization}, while paired acoustic-articulatory recordings are costly to
collect \cite{parrot2020independent}.
Speaker-independent models and adaptation methods address this challenge using vocal tract length normalization \cite{sivaraman2019unsupervised} and self-supervised speech
representations
\cite{wu2023speaker}.

The articulatory representation also determines the scope
of inversion. Electromagnetic articulography (EMA) tracks
sparse sensors on easily accessible articulators, whereas real-time magnetic resonance imaging (rt-MRI) captures midsagittal
anatomy, including pharyngeal and laryngeal regions.
Previous studies have reconstructed rt-MRI images from
speech \cite{csapo2020speaker}, while automatic segmentation
has enabled contour-based representations
\cite{ribeiro2024automatic}.
Azzouz et al. demonstrated acoustic reconstruction of
contours spanning the vocal tract from the glottis to the
lips \cite{azzouz2025reconstruction}.
Transferring these predictions to unseen speakers,
however, requires addressing geometric mismatch.

Geometric registration has been investigated through affine
mappings between EMA-derived articulatory systems
\cite{cho2024self}.
Wei and Dang used grid-defined landmarks and thin-plate
spline (TPS) warping to normalize multi-speaker EMA data
to a shared template \cite{wei2010morphological}.
Anatomical information has also been incorporated through
joint estimation of anatomy and articulation
\cite{sun2022unsupervised} and speech-conditioned rt-MRI
generation using static MRI \cite{shi2025speech}.
Here, we instead use anatomical landmarks to transfer complete MRI-derived contour predictions from a fixed AAI model to unseen speaker geometries using only single-frame anatomical calibration, without model retraining. We examine how landmark coverage and nonlinear alignment contribute under this constraint.

We estimate an affine-TPS reference-to-target mapping from primarily vertebrae and dental landmarks identified in one /u/ frame per speaker under comparable phonetic conditions. TPS provides smooth nonrigid deformation constrained by landmark correspondences. Each mapping is reused across recordings without updating the inversion model. We train on a single-speaker rt-MRI database and evaluate on eight speakers from a separate multi-speaker database across 10 structures: the upper and lower lips, tongue, soft-palate midline, pharyngeal wall, epiglottis, arytenoid cartilage, vocal folds, and upper and lower incisors. Comparing affine and affine-TPS configurations with 12 or 14 landmarks assesses the contributions of additional correspondences and nonlinear deformation.

\section{Datasets}
\label{sec:datasets}

We use two rt-MRI corpora from the ArtSpeech databases (ASD) acquired at the Centre Hospitalier
R\'egional Universitaire de Nancy: ASD2 for training the reference inversion model and ASD1 for evaluating cross-speaker adaptation.
ASD1 comprises five male and five female native French speakers, each contributing approximately 15 minutes of speech across 77 sentences, distributed over 16 sessions (15 usable for P4). The evaluated ASD1 data contain approximately 130,000 non-silent frames
\cite{isaieva2021multimodal,ribeiro2023deep}.
We evaluate P1 and P3-P9, excluding P2 because
insufficient laryngeal visibility prevents reliable evaluation of the retained structures.

ASD2 contains approximately 3.5 hours of speech from a
single native French female speaker, with about 2100
sentences across 153 acquisitions
\cite{azzouz2026acoustic}.
Her presence as P10 in ASD1 provides a shared speaker
identity across corpora \cite{ribeiro2023deep}.

Following the contour-tracking framework
in \cite{ribeiro2023deep}, each of these 10 structures is represented by 50 two-dimensional points, yielding 1000 output coordinates per frame.

\section{METHODS}

\subsection{Acoustic-to-articulatory inversion model}
\label{sec:inversion}
We use the backbone introduced by Azzouz et al. \cite{azzouz2025reconstruction}, comprising two fully connected layers followed by two bidirectional long short-term memory (Bi-LSTM) layers, each with 300 units.

ASD2 acquisitions are rigidly registered using the
static-head registration procedure of
Azzouz et al.~\cite{azzouz2026acoustic}.
Reference landmarks are identified after registration,
in the same coordinate system as the training contours
and denormalized predictions. Previous work uses local moving normalization to compensate
for slow variations in articulatory coordinates
\cite{parrot2020independent,azzouz2025reconstruction}.
For geometric adaptation, we instead use global
coordinate-wise statistics computed exclusively from the
reference training data:
\begin{equation}
\begin{aligned}
\widetilde{\mathbf{y}}_t
&=
(\mathbf{y}_t-\boldsymbol{\mu}_{\mathrm{train}})
\oslash \boldsymbol{\sigma}_{\mathrm{train}},
\\
\widehat{\mathbf{y}}^{R}_t
&=
\widehat{\widetilde{\mathbf{y}}}_t
\odot \boldsymbol{\sigma}_{\mathrm{train}}
+\boldsymbol{\mu}_{\mathrm{train}},
\end{aligned}
\label{eq:normalization}
\end{equation}
where $\oslash$ and $\odot$ denote element-wise division
and multiplication, respectively.
The same training statistics are reused during validation
and inference to recover reference-space contour
coordinates. No target-speaker contour sequences are
required to estimate the articulatory normalization
statistics.

\subsection{Geometric adaptation}
\label{sec:geometry}

For each target speaker $s$, anatomical landmark correspondences define a fixed mapping
$G_{R\rightarrow s}$
from predicted reference-space contours to the target geometry, reused across recordings.

\begin{figure}[t]
    \centering
    \includegraphics[width=\columnwidth]
        {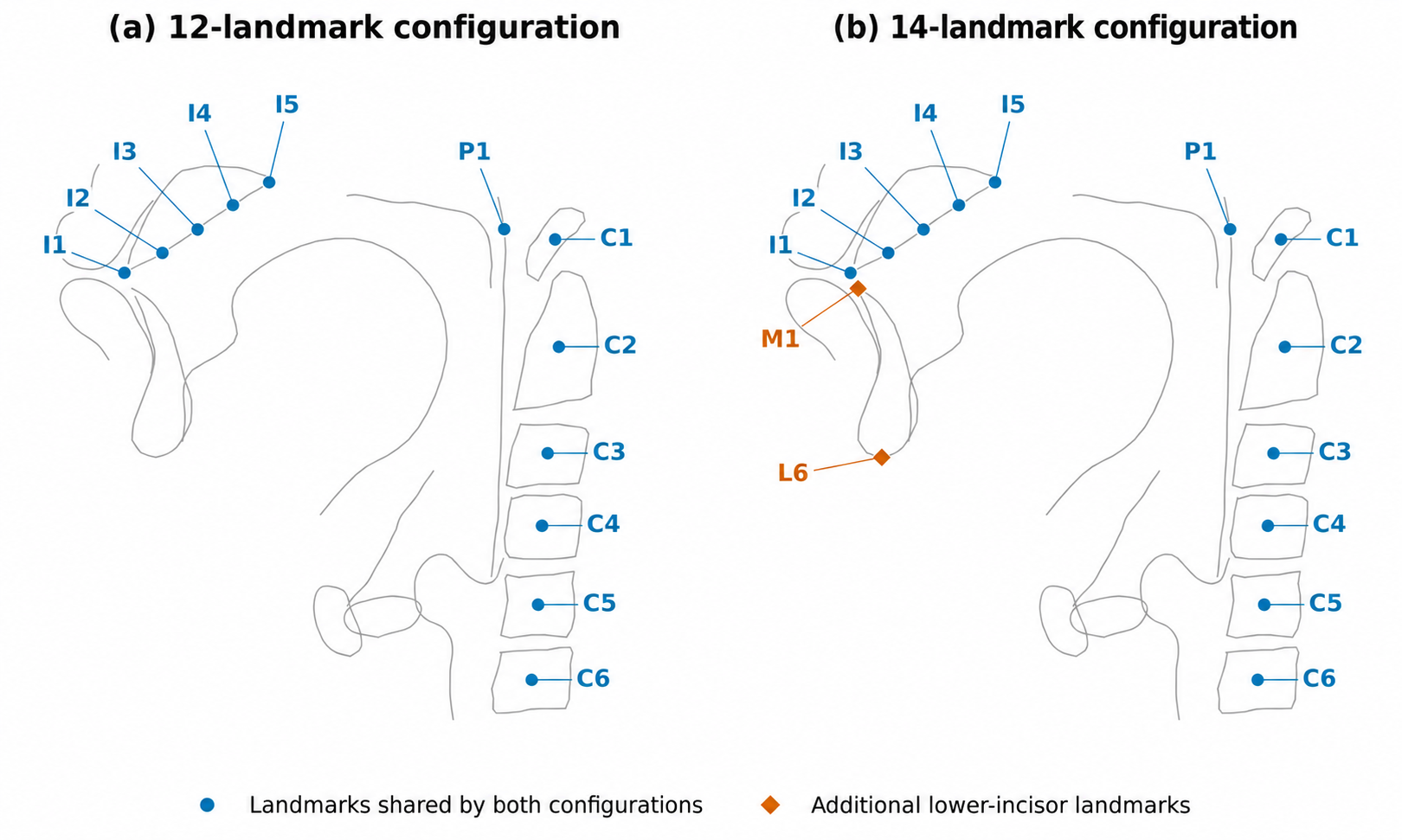}
    \caption{Anatomical landmark configurations used for geometric adaptation. The 12-landmark set (in blue) is shown together with the two additional lower-incisor landmarks, M1 and L6 (in orange), used in the 14-landmark configuration.}
    \label{fig:landmarks}
\end{figure}

\subsubsection{Anatomical landmarks}

As shown in Fig.~\ref{fig:landmarks}, the 12-landmark
configuration comprises five points sampled along the
inferior boundary of the upper-incisor and hard-palate
contour ($I1$-$I5$),
the centers of the cervical vertebrae ($C1$-$C6$),
and a posterior landmark $P1$.
The bony and dental landmarks provide anatomical anchors,
while $P1$ supplies an additional correspondence on the
pharyngeal wall.

The landmark $P1$ is defined as the intersection of the
line through $I5$ and the center of $C1$ with the
pharyngeal-wall contour.
The 14-landmark configuration additionally includes
$M1$ and $L6$, defined as the superior and inferior extrema,
respectively, of the lower-incisor contour in the midsagittal
image. These landmarks extend coverage to the lower anterior oral cavity and constrain mandibular alignment.
These landmark definitions are applied to the calibration frames described in Section~\ref{sec:experiments}.

\subsubsection{Geometric transformation}

In the proposed configuration, the 12-landmark set first estimates a global affine alignment accounting for translation, rotation, scaling, and shear, after which M1 and L6 provide two additional constraints for the subsequent nonrigid deformation:
\begin{equation}
(\mathbf{A}_{R\to s},\mathbf{b}_{R\to s})
=
\underset{\mathbf{A},\mathbf{b}}{\arg\min}
\sum_{k=1}^{12}
\lVert
\mathbf{A}\mathbf{l}_{R,k}
+\mathbf{b}-\mathbf{l}_{s,k}
\rVert_2^2 .
\label{eq:affine-fit}
\end{equation}
where $\mathbf{l}_{R,k}$ and $\mathbf{l}_{s,k}$ are the
corresponding two-dimensional landmark coordinates,
$\mathbf{A}_{R\rightarrow s}\in\mathbb{R}^{2\times2}$,
and $\mathbf{b}_{R\rightarrow s}\in\mathbb{R}^{2\times1}$.
The affine transformation acts on a reference point $\mathbf{p}\in\mathbb{R}^{2\times1}$ as
$T_{R\rightarrow s}^{\mathrm{Affine}}(\mathbf{p})
=\mathbf{A}_{R\rightarrow s}\mathbf{p}
+\mathbf{b}_{R\rightarrow s}$.

The affine transformation is then applied to all 14 reference
landmarks, including M1 and L6, giving
$\mathbf{c}_{s,k}
=T_{R\rightarrow s}^{\mathrm{Affine}}(\mathbf{l}_{R,k})$.
TPS is fitted from these affine-transformed landmarks
to their target correspondences $\mathbf{l}_{s,k}$:
\begin{equation}
T_{R\rightarrow s}^{\mathrm{TPS}}(\mathbf{q})
=\mathbf{a}_{s,0}+\mathbf{B}_s\mathbf{q}
+\sum_{k=1}^{14}\mathbf{w}_{s,k}
U\!\bigl(\lVert\mathbf{q}-\mathbf{c}_{s,k}\rVert_2\bigr).
\label{eq:tps}
\end{equation}
where $\mathbf{q},\mathbf{a}_{s,0},\mathbf{c}_{s,k},
\mathbf{w}_{s,k}\in\mathbb{R}^{2\times1}$,
$\mathbf{B}_s\in\mathbb{R}^{2\times2}$,
and $U(r)=r^2\log r$ for $r>0$, with $U(0)=0$
\cite{bookstein1989principal}. TPS is fitted by exact interpolation
with zero smoothing and a first-degree polynomial term.
The complete reference-to-target mapping is
\begin{equation}
G_{R\rightarrow s}(\mathbf{p})
=
T_{R\rightarrow s}^{\mathrm{TPS}}
\bigl(T_{R\rightarrow s}^{\mathrm{Affine}}(\mathbf{p})\bigr).
\label{eq:complete-mapping}
\end{equation}
This mapping is applied point-wise to the predicted
reference-space contours.

\subsection{Evaluation}
\label{sec:evaluation}

Following Ribeiro \cite{ribeiro2023deep}, we evaluate
contour reconstruction using the mean
point-to-closest-point distance (P2CP$_{\mathrm{mean}}$).
Unlike point-wise RMSE, this metric does not require
matching point indices across predicted and target
contours, making it suitable for cross-speaker geometric
evaluation.

Before evaluation, predicted and target contours are independently regularized using quadratic B-splines
with 20 subintervals and a regularization parameter of
0.1, then resampled to $N=50$ points.
For contours $\mathbf{u}$ and $\mathbf{v}$,
\begin{equation}
\mathrm{P2CP}_{\mathrm{mean}}(\mathbf{u},\mathbf{v})
=
\frac{1}{2N}
\sum_{i=1}^{N}
\left[
\min_j d(\mathbf{u}_i,\mathbf{v}_j)
+
\min_j d(\mathbf{v}_i,\mathbf{u}_j)
\right],
\label{eq:p2cp}
\end{equation}
where $d$ denotes the Euclidean distance.
Distances are reported in millimeters.
For each structure, mean errors are first computed over
frames within each session, then averaged equally across
sessions and speakers. Aggregate means give equal weight
to the selected structures.

Standard deviations (SDs) describe pooled frame-level errors for individual structures and pooled frame structure errors for
speaker-wise results. Means retain the equal-weight averaging described above.

\subsection{Experimental setup}
\label{sec:experiments}

The reference inversion model is trained on ASD2 using
the partitions and training settings of
Azzouz et al.~\cite{azzouz2026acoustic}, with the
10-structure output and global normalization described
in Section~3.1.
Acoustic inputs comprise 13 MFCCs and their first-
and second-order derivatives (39 dimensions), extracted
with a 25-ms window and a 10-ms hop.
MRI contours sampled every 20 ms are aligned to the
10-ms acoustic grid by inserting coordinate-wise averages
of consecutive contours \cite{azzouz2025reconstruction}.
The lowest-validation-RMSE checkpoint is kept fixed throughout adaptation.

Calibration uses one /u/ frame from `pour' in ASD2 and `pourri' for each ASD1 target speaker.
The vowel /u/ was selected because it provides a relatively constrained articulatory configuration \cite{ouni2005modeling}, without assuming identical articulation across speakers.
Although the reference speaker also produces
`pourri' in ASD1 (P10), we use the registered
ASD2 frame to match the geometry of the model predictions.
The reference speaker's recordings differ in head posture
between corpora, which can alter vocal-tract shape,
particularly in the pharyngeal region
\cite{douros2019effect}.
The words share the immediate /p/-/u/-/\textinvscr/
context, although their broader contexts differ.

Frames nearest the /u/ midpoint are selected from manually corrected TextGrid phonetic segmentation, independently of reconstruction errors.
One mapping is estimated per target speaker and reused across recordings; reference frames are excluded from evaluation.

Raw denotes unadapted predictions. A12 (Affine12) and A14 (Affine14) use 12 and 14 landmarks for affine alignment, respectively. A12+T12 (Affine12+TPS12) and A12+T14 (Affine12+TPS14) apply affine alignment with 12 landmarks, followed by TPS fitted using 12 and 14 affine-transformed reference landmarks, respectively. The 14-landmark set adds M1 and L6. A14 is the affine baseline using the same overall landmark set as A12+T14.
All configurations share predictions, calibration frames, evaluation data, and common-landmark definitions.

P2CP$_{\mathrm{mean}}$ is reported for the 10 evaluated
structures (Full10) and for a seven-structure subset
excluding the arytenoid cartilage, epiglottis, and vocal
folds (Full7). Full7 assesses whether improvements extend beyond the three laryngeal structures.
A12+T14 is compared with A14 using two-sided paired
$t$-tests at an unadjusted significance level of 0.05.
Paired observations are session means for within-speaker
tests and session-balanced speaker means for
articulator-wise and Full10/Full7 tests ($n=8$ speakers).

\section{Results}
\label{sec:results}
The ASD2-trained model achieves a mean P2CP$_{\mathrm{mean}}$
of $1.40$ mm across the 10 evaluated structures
(Table~\ref{tab:asd2-baseline}). On target speakers, the
unadapted Full10 error reaches $8.87$ mm, indicating
substantial geometric mismatch
(Table~\ref{tab:speaker-adaptation}).

A12+T14 achieves the lowest Full10 error of $3.19$ mm,
improving over A12 ($4.10$ mm), A14 ($3.69$ mm), and
A12+T12 ($3.75$ mm)
(Table~\ref{tab:geometry-adaptation}). The reduction
relative to the main affine baseline, A14, is approximately
$13.6\%$ and statistically significant ($p=0.025$).
Full7 error also decreases significantly, from $3.17$ to
$2.71$ mm ($p<0.05$).

A12+T14 yields the lowest error for six of the 10
structures and improves over A14 for eight. The largest
absolute reductions occur for the vocal folds ($6.07$ to
$4.86$ mm), pharyngeal wall ($3.02$ to $2.19$ mm), and
arytenoid cartilage ($4.34$ to $3.64$ mm). Tongue error
decreases from $4.06$ to $3.62$ mm. At the structure
level, only the arytenoid cartilage and lower incisor
show statistically significant reductions ($p<0.05$).
Epiglottis and soft-palate errors increase slightly,
indicating uneven benefits across structures.

All adaptation configurations outperform Raw for every
target speaker. A12+T14 performs best for seven of eight
speakers, with significant improvements over A14 for
these seven ($p<0.05$). For P4, however, error increases
significantly from $2.95$ to $3.17$ mm. A12+T14
speaker-wise errors range from $2.46$ to $3.95$ mm.

\begin{figure}[!t]
    \centering
    \includegraphics[width=\columnwidth]
    {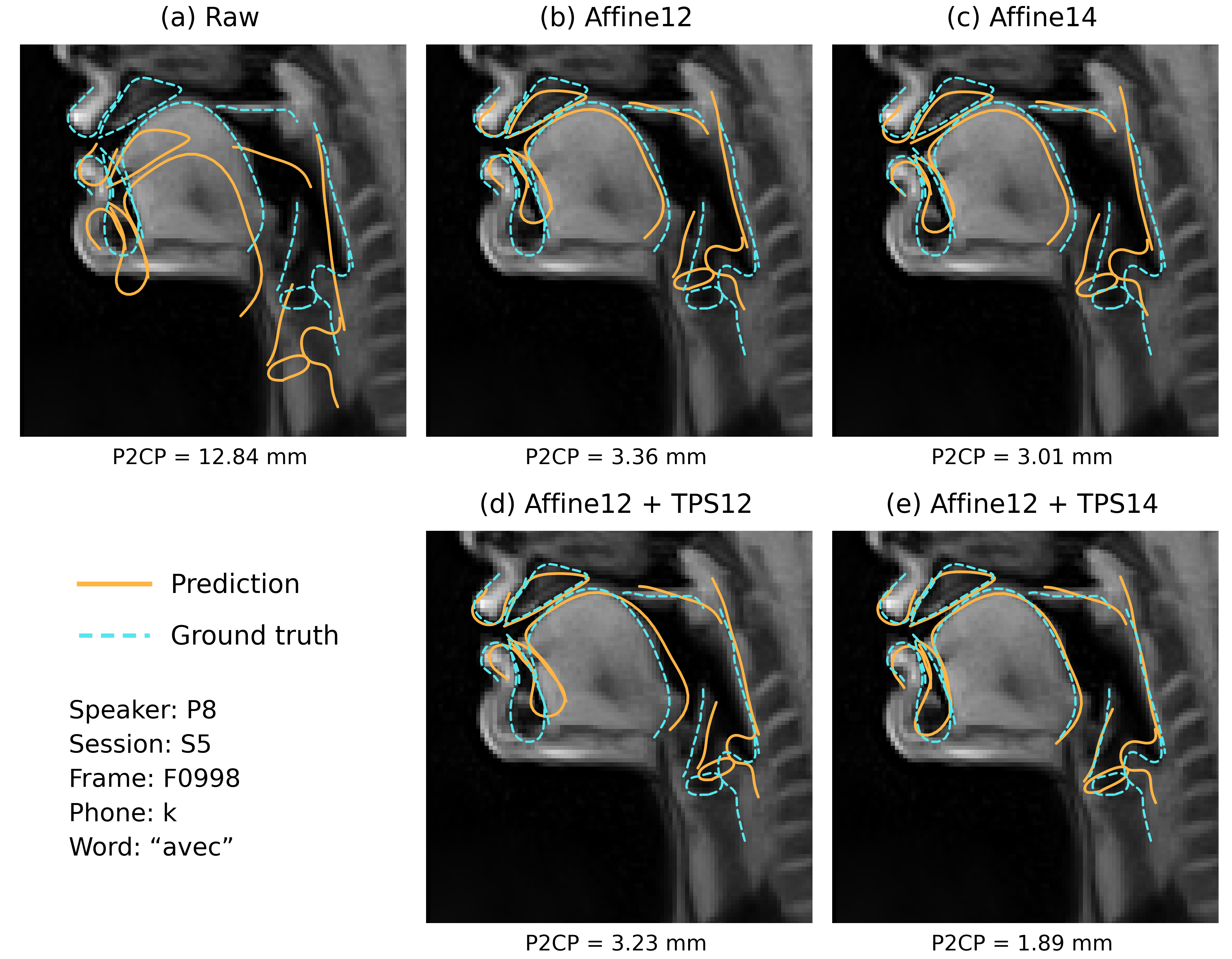}
    \caption{Adaptation example for P8  containing /k/ in \textit{avec}. Orange solid: predictions; cyan dashed: ground truth. Frame-level P2CP errors are shown below each panel.}
    \label{fig:adaptation-example}
\end{figure}

Figure~\ref{fig:adaptation-example} illustrates the
application of the fixed mapping to a P8 /k/ frame in
\textit{avec}, beyond the /u/ calibration vowel.

\begin{table}[t]
\centering
\caption{ASD2 model trained on ten articulators with global normalization. P2CP$_{\mathrm{mean}}$ in mm.}
\label{tab:asd2-baseline}
\small
\setlength{\tabcolsep}{4pt}
\renewcommand{\arraystretch}{1.08}
\begin{tabular}{lcc}
\hline
Articulator & Mean $\pm$ SD & Median \\
\hline
Arytenoid cartilage & 1.53 $\pm$ 0.76 & 1.38 \\
Epiglottis & 1.71 $\pm$ 1.18 & 1.39 \\
Lower lip & 1.29 $\pm$ 0.64 & 1.17 \\
Pharyngeal wall & 0.96 $\pm$ 0.44 & 0.85 \\
Soft-palate midline & 1.01 $\pm$ 0.55 & 0.88 \\
Tongue & 1.99 $\pm$ 0.74 & 1.84 \\
Upper lip & 1.37 $\pm$ 0.76 & 1.18 \\
Vocal folds & 1.55 $\pm$ 0.90 & 1.37 \\
Lower incisor & 1.29 $\pm$ 0.60 & 1.19 \\
Upper incisor & 1.32 $\pm$ 0.78 & 1.14 \\
\hline
Mean  & 1.40 $\pm$ 0.82 & 1.22 \\
\hline
\end{tabular}
\end{table}

\begin{table}[t]
\centering
\caption{Articulator-wise P2CP$_{\mathrm{mean}}$ (mm). A14 is the main affine baseline; A12+T14 is the proposed configuration.}
\label{tab:geometry-adaptation}
\fontsize{9}{11}\selectfont
\setlength{\tabcolsep}{2.5pt}
\renewcommand{\arraystretch}{1.18}
\newcommand{\artcell}[2]{#1\,\ensuremath{\pm}\,#2}
\begin{tabular*}{\columnwidth}{@{\extracolsep{\fill}}lcccc@{}}
\hline
Articulator & A12 & A14 & A12+T12 & A12+T14 \\
\hline
Arytenoid & \artcell{5.09}{2.33} & \artcell{4.34}{2.16} & \artcell{4.52}{2.27} & \artcell{\textbf{3.64}\textsuperscript{*}}{1.89} \\
Epiglottis & \artcell{4.94}{2.74} & \artcell{\textbf{4.29}}{2.40} & \artcell{5.00}{2.76} & \artcell{4.39}{2.43} \\
Lower lip & \artcell{3.26}{1.50} & \artcell{2.67}{1.06} & \artcell{2.91}{1.20} & \artcell{\textbf{2.53}}{1.11} \\
Pharynx & \artcell{2.87}{1.79} & \artcell{3.02}{1.91} & \artcell{2.29}{1.41} & \artcell{\textbf{2.19}}{1.27} \\
Soft palate & \artcell{\textbf{2.81}}{1.26} & \artcell{2.83}{1.42} & \artcell{2.91}{1.19} & \artcell{2.88}{1.15} \\
Tongue & \artcell{4.58}{1.61} & \artcell{4.06}{1.36} & \artcell{4.41}{1.43} & \artcell{\textbf{3.62}}{1.17} \\
Upper lip & \artcell{3.42}{1.11} & \artcell{3.89}{1.30} & \artcell{\textbf{3.15}}{1.15} & \artcell{3.21}{1.16} \\
Vocal folds & \artcell{7.54}{4.20} & \artcell{6.07}{3.64} & \artcell{6.59}{4.20} & \artcell{\textbf{4.86}}{2.87} \\
Lower incisor & \artcell{3.93}{1.57} & \artcell{2.83}{0.96} & \artcell{3.41}{1.10} & \artcell{\textbf{2.14}\textsuperscript{*}}{0.80} \\
Upper incisor & \artcell{2.56}{0.95} & \artcell{2.90}{1.03} & \artcell{\textbf{2.32}}{0.91} & \artcell{2.40}{0.98} \\
\hline
Full10 & \artcell{4.10}{2.56} & \artcell{3.69}{2.15} & \artcell{3.75}{2.39} & \artcell{\textbf{3.19}\textsuperscript{*}}{1.85} \\
Full7 & \artcell{3.35}{1.57} & \artcell{3.17}{1.42} & \artcell{3.06}{1.38} & \artcell{\textbf{2.71}\textsuperscript{*}}{1.21} \\
\hline
\end{tabular*}
\par\smallskip
\raggedright
Values: mean $\pm$ SD; bold: lowest mean.
$^{*}$Different from A14
(paired two-sided $t$-test, $n=8$ speakers, unadjusted $p<0.05$).
\end{table}

\begin{table}[t]
\centering
\caption{Speaker-wise Full10 P2CP$_{\mathrm{mean}}$ (mm). A14 is the main affine baseline; A12+T14 is the proposed configuration.}
\label{tab:speaker-adaptation}
\fontfamily{ptm}\fontsize{9}{11}\selectfont
\setlength{\tabcolsep}{0.4pt}
\renewcommand{\arraystretch}{1.15}
\newcommand{\spkcell}[2]{#1\ensuremath{\pm}#2}
\begin{tabular*}{\columnwidth}{@{\extracolsep{\fill}}lccccc@{}}
\hline
Speaker & Raw & A12 & A14 & A12+T12 & A12+T14 \\
\hline
P1 & \spkcell{9.71}{3.63} & \spkcell{5.85}{3.41} & \spkcell{5.33}{2.94} & \spkcell{5.34}{3.51} & \spkcell{\textbf{3.95}\textsuperscript{*}}{2.31} \\
P3 & \spkcell{9.69}{4.26} & \spkcell{5.06}{3.38} & \spkcell{4.42}{2.49} & \spkcell{4.31}{2.89} & \spkcell{\textbf{3.46}\textsuperscript{*}}{1.76} \\
P4 & \spkcell{8.14}{4.26} & \spkcell{3.28}{1.64} & \spkcell{\textbf{2.95}}{1.36} & \spkcell{3.06}{1.93} & \spkcell{3.17\textsuperscript{*}}{2.10} \\
P5 & \spkcell{4.42}{2.12} & \spkcell{3.13}{1.36} & \spkcell{2.93}{1.21} & \spkcell{2.64}{1.03} & \spkcell{\textbf{2.46}\textsuperscript{*}}{1.00} \\
P6 & \spkcell{7.79}{3.37} & \spkcell{3.41}{1.97} & \spkcell{2.94}{1.58} & \spkcell{3.32}{1.69} & \spkcell{\textbf{2.82}\textsuperscript{*}}{1.36} \\
P7 & \spkcell{11.23}{4.55} & \spkcell{3.41}{2.27} & \spkcell{3.34}{2.26} & \spkcell{3.69}{2.15} & \spkcell{\textbf{3.17}\textsuperscript{*}}{2.14} \\
P8 & \spkcell{14.22}{5.27} & \spkcell{3.87}{1.99} & \spkcell{3.79}{1.84} & \spkcell{4.41}{2.00} & \spkcell{\textbf{3.24}\textsuperscript{*}}{1.44} \\
P9 & \spkcell{5.74}{2.01} & \spkcell{4.80}{2.14} & \spkcell{3.84}{1.67} & \spkcell{3.25}{1.85} & \spkcell{\textbf{3.22}\textsuperscript{*}}{1.89} \\
\hline
All & \spkcell{8.87}{4.81} & \spkcell{4.10}{2.56} & \spkcell{3.69}{2.15} & \spkcell{3.75}{2.39} & \spkcell{\textbf{3.19}\textsuperscript{*}}{1.85} \\
\hline
\end{tabular*}
\par\smallskip
\raggedright
Values: mean $\pm$ SD; bold: lowest mean.
$^{*}$Different from A14
(paired two-sided $t$-test; session means within speakers, speaker-level for All; unadjusted $p<0.05$).
\end{table}

\section{Discussion}
These comparisons assess the contributions of anatomical correspondences and nonlinear alignment. A14 improves over A12 by adding two landmarks, whereas A12+T12 improves aggregate performance with the landmark set unchanged. A12+T14 performs best overall, suggesting that the additional lower-incisor correspondences help constrain nonrigid adaptation. Tongue error decreases by 10.8\% relative to A14, although this reduction is not statistically significant.

Residual laryngeal errors may reflect anatomical differences, including gender-related oral-pharyngeal proportions and laryngeal position
\cite{fitch1999morphology,mirjalili2012vertebral}, whose contributions are not isolated here. Lip variability and the deterioration for P4 further indicate that benefits are not uniform. A fixed mapping may only partially capture speaker-specific articulation \cite{serrurier2019characterization} and pose differences across unregistered ASD1 recordings\cite{douros2019effect}. Sensitivity to calibration-frame selection and landmark annotation remains to be quantified.

\section{Conclusion}
Single-frame anatomical adaptation transfers vocal-tract contours from a fixed AAI model to unseen speakers without retraining. Across eight speakers, Affine12+TPS14 reduces Full10 error from 3.69 mm with Affine14 to 3.19 mm, supporting landmark coverage and nonlinear alignment. Future work will extend this framework to multi-speaker training toward speaker-independent AAI, while assessing calibration robustness and residual laryngeal and lip errors.

\section{Acknowledgments}
This research was supported by the French ANR project ARTANY, CPER IT2MP, Région Lorraine and FEDER, and financed by Lorraine Université d’Excellence (LUE). The RT-MRI data were acquired on a platform of the FLI network.
\bibliographystyle{IEEEbib}
\bibliography{refs}

\end{document}